\documentclass{aa}  
\usepackage{orcidlink}
\usepackage{natbib}
\bibpunct{(}{)}{;}{a}{}{,} 

\usepackage{graphicx}
\usepackage{txfonts}
\usepackage{color}
\usepackage{threeparttable}
\begin{document}

\title{Correlated spectro-polarimetric study along the Z track in XTE~J1701--462 puts constraints on its coronal geometry}

\author{
Wei Yu\inst{1,2,3} \orcidlink{0000-0002-3229-2453}
\and Qingcui Bu\inst{4}\thanks{bu@ccnu.edu.cn} \orcidlink{0000-0001-5238-3988}
\and Victor Doroshenko\inst{3} \orcidlink{0000-0001-8162-1105}
\and Lorenzo Ducci\inst{3,5} \orcidlink{0000-0002-9989-538X}
\and Long Ji\inst{6} \orcidlink{0000-0001-9599-7285}
\and Wenda Zhang\inst{7} \orcidlink{0000-0003-1702-4917}
\and Andrea Santangelo\inst{3} \orcidlink{0000-0003-4187-9560}
\and Shuangnan Zhang\inst{1,2} \orcidlink{0000-0001-5586-1017}
\and Anand Waghmare\inst{3}
\and Mingyu Ge\inst{1} \orcidlink{0000-0002-3776-4536}
\and Yue Huang\inst{1} \orcidlink{0000-0002-3515-9500}
\and Hexin Liu\inst{1} \orcidlink{0000-0002-8032-7024}
\and Lian Tao\inst{1} \orcidlink{0000-0002-2705-4338}
\and Zixu Yang\inst{8} \orcidlink{0000-0003-1718-8487}
\and Liang Zhang\inst{1} \orcidlink{0000-0003-4498-9925}
\and Jinlu Qu\inst{1,2} \orcidlink{0000-0002-9796-2585}
}

\institute{
Key Laboratory of Particle Astrophysics, Institute of High Energy Physics, Chinese Academy of Sciences, Beijing 100049, China
\and University of Chinese Academy of Sciences, Chinese Academy of Sciences, Beijing 100049, China
\and Institut f\"ur Astronomie und Astrophysik, Kepler Center for Astro and Particle Physics, Eberhard Karls Universit\"at, Sand 1, 72076 T\"ubingen, Germany
\and Institute of Astrophysics,
Central China Normal University,
Luoyu Road 152, Wuhan 433079, P.R. China
\and ISDC Data Center for Astrophysics, Universit\'e de Gen\`eve, 16 chemin d'\'Ecogia, 1290 Versoix, Switzerland
\and School of Physics and Astronomy, Sun Yat-sen University, Zhuhai 519082, China
\and National Astronomical Observatories, Chinese Academy of Sciences, A20 Datun Road, Beijing 100012, China
\and School of Physics and Optoelectronic Engineering, Shandong University of Technology, Zibo 255000, China
}

 
  \abstract
   {In September 2022, the transient neutron star low-mass X-ray binary XTE J1701--462 went into a new outburst.}
   {The objective of this work is to examine the evolution of the accretion geometry of XTE J1701--462 by studying the spectro-polarimetric properties along the Z track of this source. The simultaneous observations archived by the \emph{Insight}-Hard X-ray Modulation Telescope (HXMT) and the Imaging X-ray Polarimetry Explorer (\emph{IXPE}) give us the opportunity to do this.}
   {We present a comprehensive X-ray spectro-polarimetric analysis of XTE J1701--462, using simultaneous observations from \emph{IXPE}, \emph{Insight}-HXMT, and \emph{NuSTAR}. For \emph{IXPE} observations, two methods were employed to measure the polarization: a model-independent measurement with \textsc{PCUBE} and a model-dependent polarization-spectral analysis with \textsc{XSPEC}. The corresponding spectra from \emph{Insight}-HXMT and \emph{NuSTAR} were studied with two configurations that correspond to the eastern model and western model, respectively.}
   {Significant polarization characteristics are detected in XTE J1701--462. The polarization degree shows a decreasing trend along the Z track, reducing from $(4.84\pm0.37)\%$ to $(3.76\pm0.43)\%$ on the horizontal branch and dropping to $\leq 1\%$ on the normal branch. The simultaneous spectral analysis from \emph{Insight}-HXMT and \emph{NuSTAR} suggests that the evolution of the PD is closely linked to changes in the covering factor of the Comptonized corona along the Z track.}



\titlerunning{Correlated spectro-polarimetric study along the Z track in XTE~J1701--462 puts constraints on its coronal geometry}
\authorrunning{Wei Yu et al.}
\maketitle

\section{Introduction}

Weakly magnetized neutron stars in low-mass X-ray binaries
(NS-LMXBs) are believed to accrete mass via Roche-lobe overflow from a low-mass
stellar companion. These sources show intensity and spectral variations on timescales ranging from hours to months, and are historically divided into two classes, Z and atoll sources. The classification is based on the shape of their tracks in the X-ray color-color diagram (CCD), or in the hardness-intensity diagram (HID) \citep{1989A&A...225...79H}. The Z sources trace a Z-shaped track with three branches: a horizontal branch (HB), a normal branch (NB), and a flaring branch (FB). The Z sources are further classified into two subcategories based on the shape and orientation of their branches. The Cyg-like Z sources have prominent HBs and weak FBs, and vice versa for the Sco-like Z sources.

XTE J1701--462 is a NS-LMXB discovered with the All Sky Monitor (ASM) of the Rossi X-ray Timing Explorer (RXTE) in January 2006 \citep{2006ATel..696....1R}.
Among the NS-LMXBs, XTE J1701--462 can be regarded as a rather unique object that is the first source to show both Z-like and Atoll-like behavior during one single outburst \citep{2007ApJ...656..420H,2009ApJ...696.1257L,2010ApJ...719..201H}. In the first 10 weeks of its outburst, XTE J1701--462 evolved from a Cyg-like source into a Sco-like Z one, and during the decay it showed an Atoll-like behavior \citep{2007ApJ...656..420H,2009ApJ...696.1257L, bu2015correlations}. The orbital inclination of the system is lower than $75^\circ$ \citep{2009ApJ...696.1257L}. In September 2022, XTE J1701--462 experienced a new outburst, the second one since its discovery in 2006 \citep{2022ATel15592....1I}. 

The X-ray emission of NS-LMXBs can be generally described by models that include a soft component, either a single-temperature blackbody emission from the neutron star (NS) surface or a multicolor blackbody emission from the accretion disk, and a hard component resulting from Comptonization of soft photons scattered by the hot electron plasma in the corona. Based on the choices of the thermal and Comptonized components, there are two classical models, the ``eastern model'' and the ``western model.'' In the eastern model, the thermal component is described by a multicolor disk, while the Comptonized component is described by a weakly Comptonized blackbody that may originate from the boundary layer (BL) between the disk and the NS \citep{1977MNRAS.178..195P,1988AdSpR...8b.135S,2001ApJ...547..355P} or the spreading layer (SL) near the surface of the NS \citep{1999AstL...25..269I,2006MNRAS.369.2036S}. 
In the western model, the thermal component is a single-temperature blackbody from the BL and the Comptonized emission is from the disk \citep{1988ApJ...324..363W, 2000A&A...360L..35P,2007A&A...471L..17P,2000ApJ...544L.119D,2001ApJ...554...49D,2012ApJ...752L..34C,2019MNRAS.484.3004W}.

Because of the strong degeneracy found in the spectral models \citep{2001AdSpR..28..307B}, spectroscopy alone is usually insufficient to give a full picture of the accretion geometry. In addition to spectroscopy, both timing and polarization analysis can offer extra perspectives on its examination. For instance, the Fourier-frequency resolved spectroscopy analysis of a sample of atoll and Z sources has provided extra support for the eastern model, based on the observation that the Fourier-frequency resolved spectrum closely resembles the SL spectrum at the frequencies of quasiperiodic oscillations \citep{2003A&A...410..217G,2006A&A...453..253R,2013MNRAS.434.2355R}. Besides, X-ray polarimetry offers a unique perspective on the geometry of the system, while eliminating the degeneracy left by spectroscopy. In particular, Compton scattering yields a polarization signal that is highly sensitive to the geometry of the scattering material \citep{2022MNRAS.514.2561G,2023ApJ...943..129C}. Various geometries and viewing angles will result in markedly different polarization degrees (PDs) and polarization angles (PAs).  

The Imaging X-ray Polarimetry Explorer (\emph{IXPE}) operates in the 2-8 keV band, providing a unique opportunity to investigate the accretion geometric evolution through X-ray polarimetry and constraining the coronal geometry of NS-LMXBs. To date, \emph{IXPE} has targeted a variety of NS-LMXBs, including the Z sources Cyg X--2 \citep{2023MNRAS.519.3681F} and GX 5--1 \citep{2024A&A...684A.137F}, the Z-atoll transient source XTE J1701--462 \citep{2023MNRAS.525.4657J,2023A&A...674L..10C}, and two atoll sources, GS 1826--238 \citep{2023ApJ...943..129C} and GX 9+9 \citep{2023MNRAS.521L..74C,2023A&A...676A..20U}.

In this work, we conduct a spectro-polarimetric analysis of the NS-LMXB XTE J1701--462, using observations from \emph{IXPE}, \emph{NuSTAR}, and \emph{Insight}-HXMT. The paper is structured as follows. In Section~\ref{sec2}, we describe the observations and data reduction. In Section~\ref{sec3}, we report on the analysis of the spectral and polarimetric data. Section~\ref{sec4} is devoted to the discussion of the results. Finally, we summarize our conclusions in Section~\ref{sec5}.

\section{Observations and data reduction}\label{sec2}

\subsection{IXPE}

\emph{IXPE} observed the NS-LMXB XTE J1701--462 on September 29-30, 2022 (Epoch1), and October 8-9, 2022 (Epoch2), with net exposure times of 46.2 ks and 46.4 ks, respectively. We analyzed the \emph{IXPE} data using both the Python-based software \texttt{IXPEobssim} v30.2.21 \citep{2022SoftX..1901194B} and \textsc{XSPEC}. The source region is defined as a circular area with a radius of $60''$, while the background region is defined as an annular area with an inner radius of $180''$ and an outer radius of $240''$.

The polarization parameters (PD and PA) were extracted in the energy band 2--8 keV using the model-independent \textsc{PCUBE} algorithm of \texttt{XPBIN} tool \citep{2015APh....68...45K}. We also performed a spectro-polarimetric model-dependent fit of the data using \textsc{XSPEC}. Source and background spectra corresponding to the Stokes parameters I, Q, and U for all DUs were extracted using \texttt{XSELECT}. The latest \emph{IXPE} response matrices (v12) were used in the spectral fitting.

\subsection{Insight-\emph{HXMT}}

\emph{Insight}-HXMT observed the source from September 7 to October 2, 2022, which fully overlaps with Epoch1 of the \emph{IXPE} observations. There are three telescopes on board \emph{Insight}-HXMT: the high-energy X-ray telescope \citep[HE: 20-250 keV;][]{2020SCPMA..6349503L}, the medium-energy X-ray telescope \citep[ME: 5--30 keV;][]{2020SCPMA..6349504C}, and the low-energy X-ray telescope \citep[LE: 1--15 keV;][]{2020SCPMA..6349505C}.

The data were processed using the \texttt{hpipeline} under \emph{Insight}-HXMT Data Analysis Software (HXMTDAS) version 2.05. The data were filtered using the criteria recommended by the \emph{Insight}-HXMT team: a pointing offset angle smaller than $0.04^\circ$; an elevation angle greater than $10^\circ$; a geomagnetic cutoff rigidity value larger than 8; and data usage at least 300 s before and after the South Atlantic Anomaly passage. To avoid potential contamination from bright Earth and nearby sources, only small fields of view were applied. Given that the spectrum above 30 keV is dominated by the background for this source, we further selected the 2--30 keV energy range for our spectral fitting.

\subsection{NuSTAR}

Since \emph{Insight}-HXMT lacks the simultaneous observation of Epoch2, we included the \emph{NuSTAR} observation in our spectral analysis. \emph{NuSTAR} observed the source on October 8, 2022 (ObsID 90801325002), with a net exposure of 12.2 ks, which overlaps with Epoch2 of \emph{IXPE}. We processed the \emph{NuSTAR} data using v.1.9.7 of the \texttt{NuSTARDAS} pipeline with the latest \textsc{CALDB} files. The spectra were extracted from a circular region of the radius $60''$ centered on the source location. The background was estimated from a blank region on the detector furthest from the source location to avoid source photons. The spectra and light curves were extracted using the \texttt{nuproduct} task. We re-binned the spectra with 50 counts per bin by using the \texttt{ftgrouppha} task. Similar to \emph{Insight}-HXMT, the selected energy range is 3--30 keV. 

\begin{figure}
	\centering\includegraphics[width=\columnwidth]{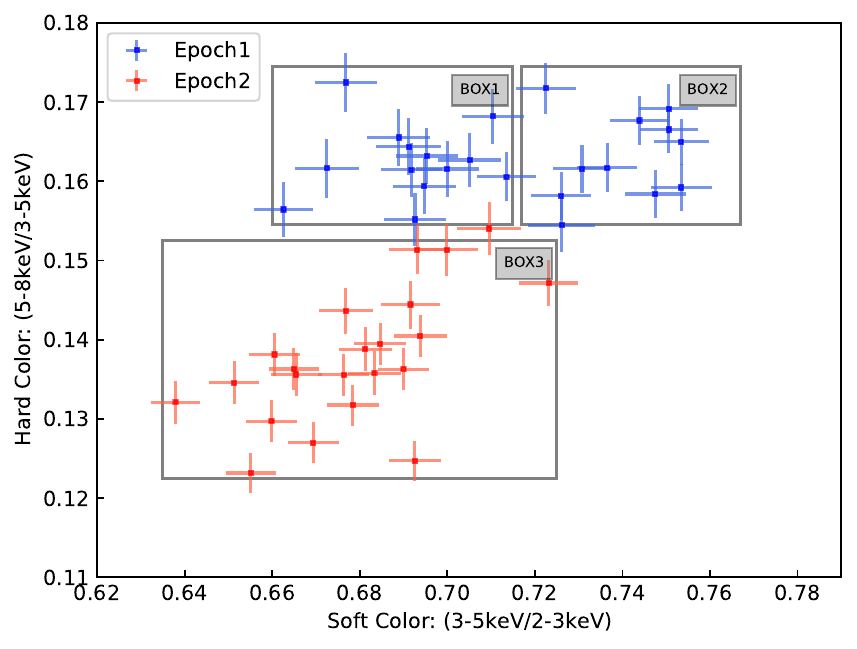}
    \caption{\emph{IXPE} colour--colour diagram of XTE J1701--462. The hard and soft colours are defined as the count rate ratios 5–-8 keV/3-–5 keV and 3–-5 keV/2-–3 keV, respectively. The blue points correspond to Epoch1 in the HB, while the orange points correspond to Epoch2 in the NB. }
    \label{fig1}
\end{figure}

\begin{figure}
	\centering\includegraphics[width=\columnwidth]{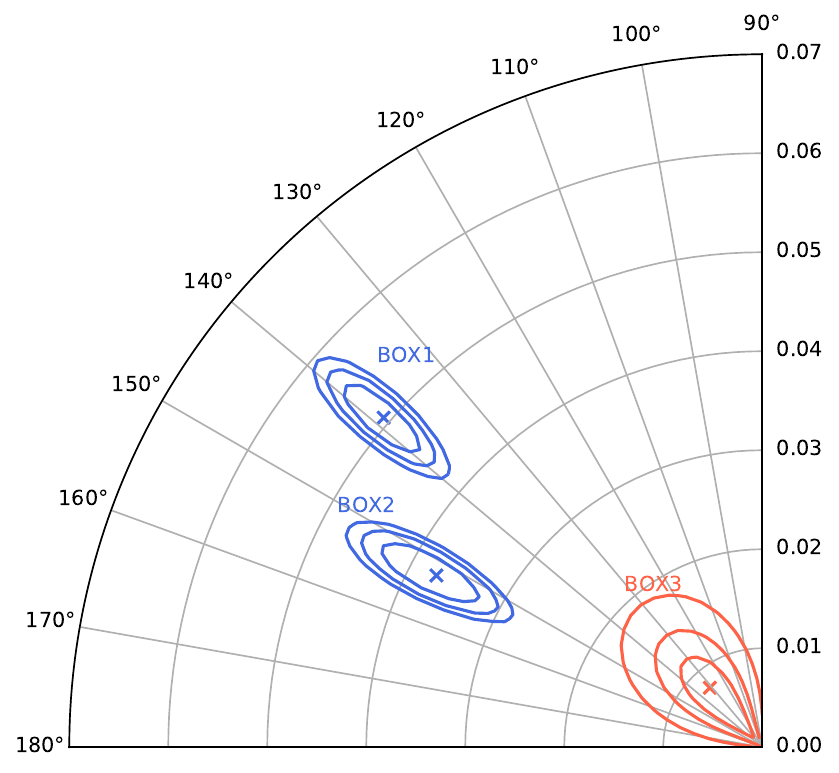}	
    \caption{Contour plot of PD and PA at 68.27, 95.45, and 99.73 percent confidence levels in the 2--8 keV band obtained with \textsc{XSPEC}. The polarization-spectral fitting was performed using model: tbabs*polconst(diskbb+thcomp*bbodyrad+gaussian). The parameters are shown in Table~\ref{table1}}
    \label{fig2}
\end{figure}

\renewcommand\arraystretch{1.5}
\begin{table*}[htbp]
	\centering
	\caption{Polarimetric parameters measured with \textsc{PCUBE} and \textsc{XSPEC}, respectively.  The polarization-spectral fitting was performed using model: tbabs*polconst(diskbb+thcomp*bbodyrad+gaussian). The uncertainties are given at a 1$\sigma$ confidence level.}
	\label{table1}
        \setlength\tabcolsep{15pt}
	\begin{tabular}{cccc}
            \multicolumn{4}{c}{PCUBE} \\
		\hline
		Parameter   & BOX1 & BOX2  & BOX3  \\    
            \hline
            Location    & HB   & HB   & NB \\
            \hline
            PD (\%)    & $4.98\pm0.44$ & $3.79\pm0.54$  & $0.9\pm0.3$   \\
            PA ($^{\circ}$) & $139.72\pm2.58$ & $149.20\pm4.29$ & $130.0\pm13.6$ \\
            \hline
            \multicolumn{4}{c}{XSPEC} \\
            \hline
            Parameter   & BOX1 & BOX2 & BOX3   \\ 
            \hline
            Location    & HB   & HB   & NB \\
            \hline
            PD (\%)    & $5.07\pm0.32$ &  $3.72\pm0.35$ &  $0.8\pm0.3$  \\
            PA ($^{\circ}$) & $136.42\pm3.44$ & $152.31\pm2.43$ & $135.3\pm12.6$ \\
            \hline
	\end{tabular}

\end{table*}

\section{Results}\label{sec3}

\subsection{Polarimetric analysis}

The CCD of the source for the \emph{IXPE} observations is shown in Fig.~\ref{fig1}. We define the soft color as the ratio of count rates in the 3--5 keV to the 2--3 keV bands, and the hard color as the ratio of count rates in the 5--8 keV to the 3--5 keV energy bands. During Epoch1, the source is observed in the HB. During Epoch2, it moves to the NB \citep{2023MNRAS.525.4657J, 2023A&A...674L..10C}.

Polarization measured from the model-independent method \textsc{PCUBE} gives a significant PD of $4.5\pm0.3\%$ and a PA of $143.5\pm2.3^{\circ}$ for HB, but a much weaker PD of $0.9\pm0.3\%$ for NB, which is less than the minimum detectable polarization at the 99\% confidence level. These results are in agreement with those reported by \citet{2023MNRAS.525.4657J} and \citet{2023A&A...674L..10C}.

\begin{figure*}
	\centering\includegraphics[width=\columnwidth]{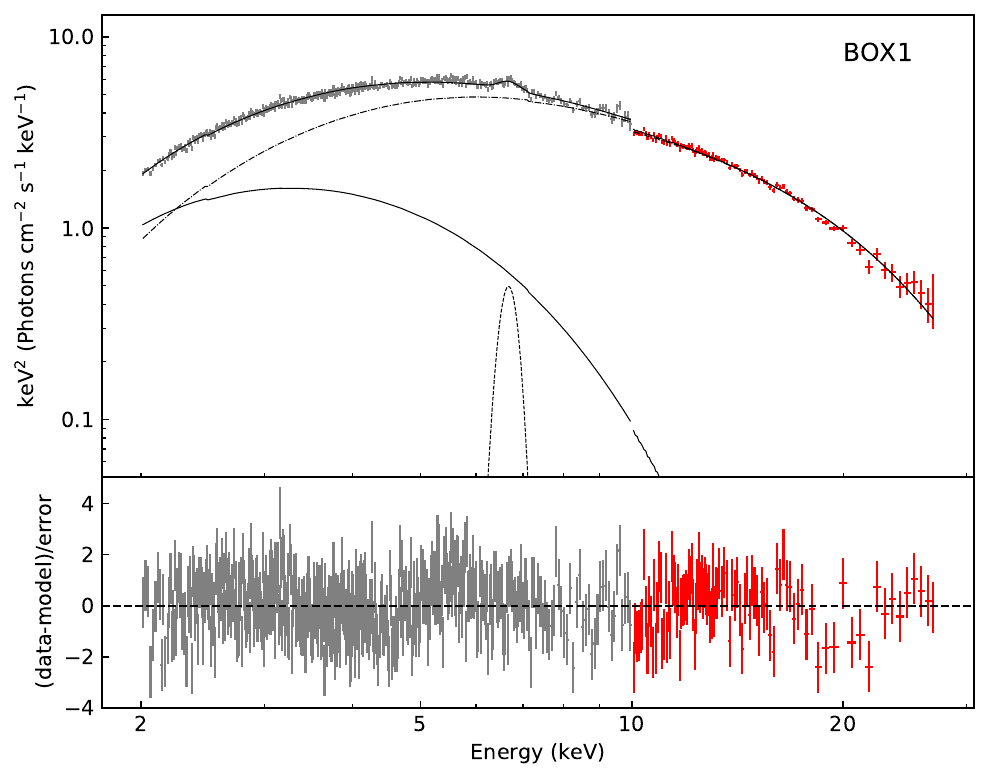}
	\centering\includegraphics[width=\columnwidth]{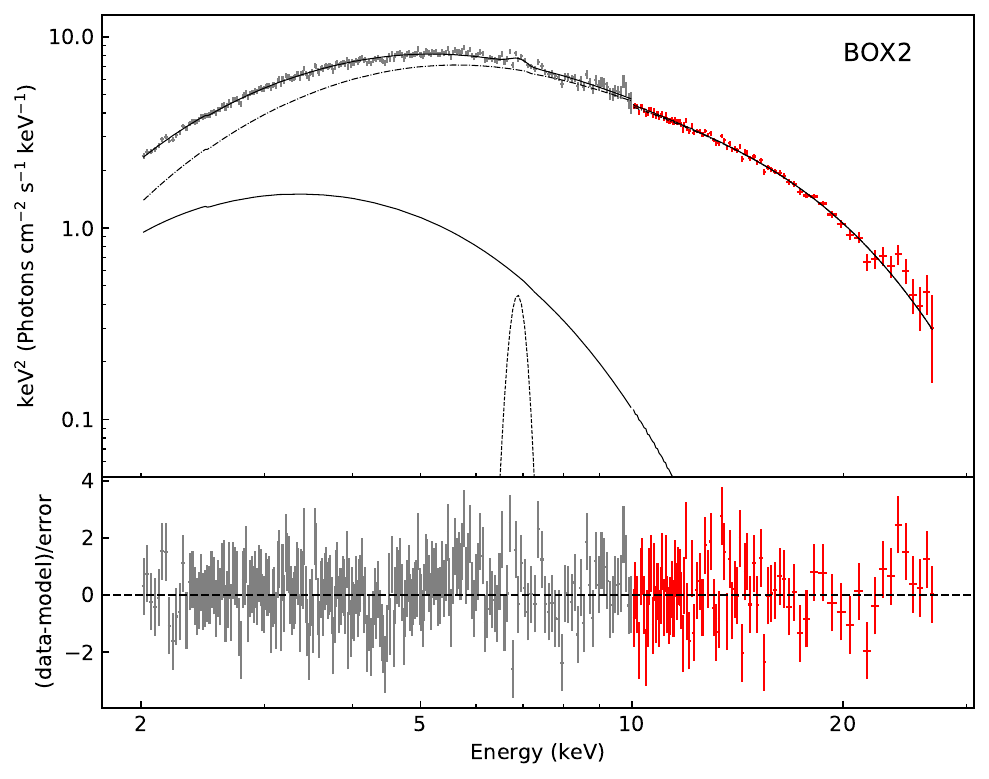}	
    \caption{\emph{Insight}-HXMT spectrum fit with Model 1. The gray and red points represent LE and ME, respectively. The full model is shown with a thick, solid line, the thermal emission \texttt{diskbb} with a dashed line, and the Comptonization component \texttt{thcomp} with a dash-dotted line.}
    \label{fig3}
\end{figure*}

\begin{figure}
    \centering\includegraphics[width=\columnwidth]{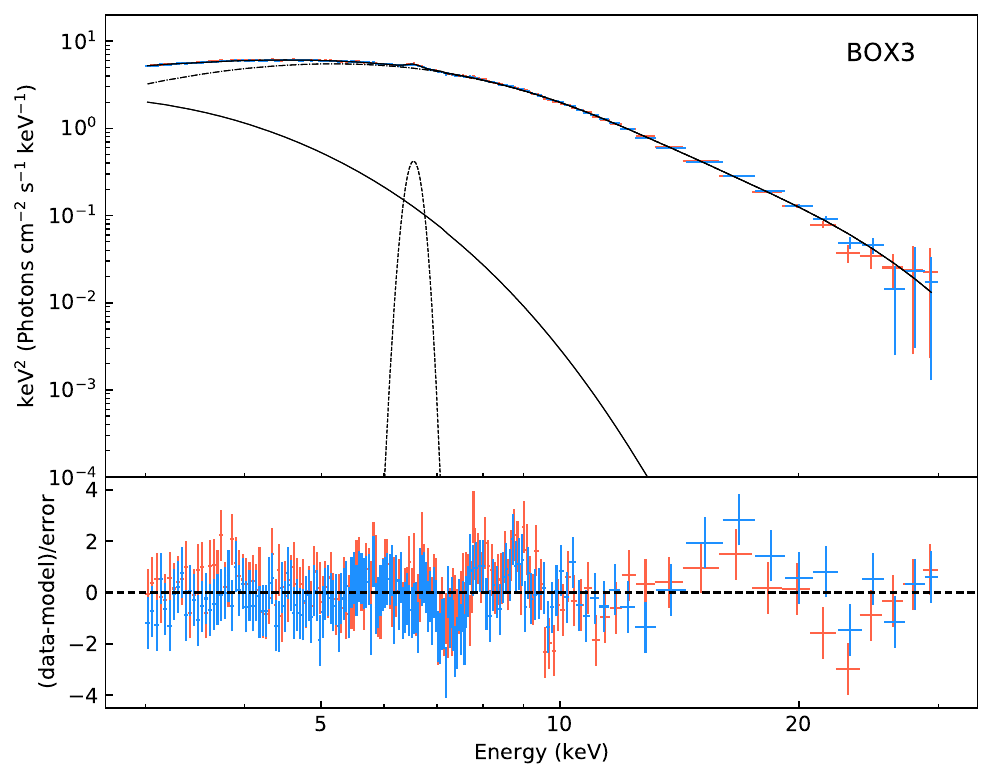}
    \caption{\emph{NuSTAR} spectrum fit with Model 1. The blue and orange points represent FPMA and FPMB, respectively. The full model is shown with a thick, solid line, the thermal emission \texttt{diskbb} with a dashed line, and the Comptonization component \texttt{thcomp} with a dash-dotted line.}
    \label{fig4}
\end{figure}

To investigate the evolution of the polarization characteristics along the Z track, we further divided Epoch1 into two segments, BOX1 and BOX2, and studied their polarization characteristics separately. The total exposure time of BOX1 and BOX2 is approximately equal. Given that the PD for Epoch2 is below the minimum detectable limit, we kept the original selection for it and defined it as BOX3. Due to the low statistics, a further study on the energy-dependent polarization of BOX1 and BOX2 is not possible.

We first used the \textsc{PCUBE} method to measure the PD and PA of the three boxes. For BOX1 and BOX2, the PDs are $4.98\pm0.44\%$ and $3.79\pm0.54\%$, with corresponding PAs of $139.72\pm2.58^{\circ}$ and $149.20\pm4.29^{\circ}$, respectively. The results are listed in Table~\ref{table1}. The results show that as the source evolves through HB, the PD decreases and the PA increases, and when the source evolves to the NB, both the PD and the PA decrease. 

We further cross-checked the results by fitting the Stokes I, Q, and U spectra with \textsc{XSPEC}. We applied the model \texttt{tbabs*polconst(diskbb+thcomp*bbodyrad)} for the broadband spectral analysis. In this model, the \texttt{polconst} component carries the polarimetric information, assuming a constant PD and PA across the entire spectrum. The \texttt{thComp} component describes the Comptonized emission from hot electron plasma, while the \texttt{diskbb} component represents emission from a standard accretion disk. Limited by the statistics and energy band of the \emph{IXPE}, the parameters were fixed to the values that are given from the spectral fitting of the \emph{Insight}-HXMT observation (see Sec.~\ref{sec3.2} for more details), leaving only the normalization and polarization parameters to vary freely.

The PD and PA obtained from the polarization-spectral fitting are listed in Table~\ref{table1}. These polarization parameters are in great agreement with the results given by the \textsc{PCUBE} method. Fig.~\ref{fig2} shows the two-dimensional contour plots for PD and PA. We also attempted to measure the polarization information of individual spectral components, but due to the low statistics of the data in each box it was not possible to constrain the polarization information for each component.

\subsection{Spectral properties}\label{sec3.2}

\renewcommand\arraystretch{1.5}
\begin{table}[htbp]
	\centering
	\caption{Best-fitting spectral parameters of Model 1. The data of BOX1 and BOX2 are from \emph{Insight}-HXMT, while BOX3 is from \emph{NuSTAR}. Uncertainties are given at a 90\% confidence level for one parameter of interest.}
	\label{table1}
        \setlength\tabcolsep{5pt}
	\begin{tabular}{cccc}
 		\hline
            \multicolumn{4}{c}{Model 1 : tbabs*(diskbb+thcomp*bbodyrad+gaussian)} \\

            \hline
		Parameters   & BOX1 & BOX2  &BOX3  \\
		\hline
            \multicolumn{4}{c}{tbabs} \\
            $N_{\rm{H}}$ ($10^{22}{\rm{cm}}^{-2}$)   &   & $1.96^*$ &    \\
            \hline
            \multicolumn{4}{c}{diskbb} \\
            $kT_{\rm{in}}$ (keV)   & $1.13_{-0.06}^{+0.05}$&$1.17_{-0.06}^{+0.06}$ &$0.71_{-0.02}^{+0.02}$   \\
            $R_{\rm{in}}$ $(\rm{km})^a$  & $16.1_{-0.9}^{+1.2}$& $14.5_{-0.2}^{+0.3}$ &  $59.2_{-2.5}^{+2.3}$  \\
            \hline
            \multicolumn{4}{c}{thComp} \\
            $\Gamma$  &$2.16_{-0.04}^{+0.09}$ & $1.68_{-0.14}^{+0.13}$ & $1.18_{-0.03}^{+0.05}$  \\
            $kT_{\rm{e}}$ (keV)   & $3.53_{-0.07}^{+0.19}$ & $2.91_{-0.04}^{+0.05}$ & $2.52_{-0.05}^{+0.05}$ \\
            $f$    & $0.70_{-0.02}^{+0.05}$ & $0.39_{-0.04}^{+0.04}$ & $0.04_{-0.01}^{+0.01}$ \\
            \hline
            \multicolumn{4}{c}{bbodyrad} \\
            $kT_{\rm{bb}}$ (keV)   &$1.15_{-0.01}^{+0.02}$ & $1.18_{-0.03}^{+0.04}$ & $1.25_{-0.01}^{+0.01}$  \\
            $R_{\rm{bb}}$ $(\rm{km})^a$ & $16.2_{-0.7}^{+0.4}$ & $18.9_{-0.6}^{+0.4}$ & $15.2_{-0.2}^{+0.2}$ \\
            \hline
            
            \multicolumn{4}{c}{gaussian} \\
            LineE   & $6.66_{-0.03}^{+0.08}$ & $6.86_{-0.01}^{+0.02}$ & $6.53_{-0.02}^{+0.04}$ \\
            $\sigma$    & $0.20_{-0.05}^{+0.02}$ & $0.18_{-0.09}^{+0.08}$ & $0.13_{-0.08}^{+0.05}$ \\
            norm ($10^{-3}$)   & $5.81_{-1.63}^{+0.61}$ & $4.46_{-1.34}^{+1.24}$ & $3.28_{-0.59}^{+0.58}$ \\
            \hline
            \multicolumn{4}{c}{2--8 keV flux ratios} \\
            $F_{\rm{dikbb}}/F_{\rm{tot}}$   & $22 \%$ & $17 \%$  & $12 \%$  \\
            $F_{\rm{thcomp}}/F_{\rm{tot}}$ & $78 \%$& $83 \%$ & $88 \%$  \\
            \hline
            \multicolumn{4}{c}{2--8 keV photon flux ratios} \\
            $F_{\rm{diskbb}}/F_{\rm{tot}}$   & $ 29\%$ & $ 24\%$  & $16 \%$  \\
            $F_{\rm{thcomp}}/F_{\rm{tot}}$ & $ 71\%$& $ 76\%$ & $84 \%$  \\
            \hline
            $\chi^2$/d.o.f.   & $1216/1233$ & $1201/1233$ & $1331/1348$ \\
            \hline
	\end{tabular}
    \begin{tablenotes}
    \footnotesize
    \item{(a)}  Radia were determined for the distance of 7 kpc and inclination of 70 degrees.
    \end{tablenotes}
\end{table}

We performed detailed spectral fitting on the three boxes using the data from \emph{Insight}-HXMT and \emph{NuSTAR}. A Comptonized component \citep[\texttt{thcomp;}][]{2020MNRAS.492.5234Z} plus either a multicolor disk blackbody (\texttt{diskbb}; hereafter Model 1) or a single-temperature blackbody (\texttt{bbodyrad}; hereafter Model 2) were applied to fit the 2--30 keV spectra. For all the spectra, a neutral Galactic absorption component, modeled with \texttt{TBabs}, was added \citep{2000ApJ...542..914W}. We adopted the abundances in \citet{2000ApJ...542..914W} as appropriate for absorption by the Galactic interstellar medium and adopted the recommended cross-sections of \citet{1996ApJ...465..487V}.

In Model 1, the seed photons for Comptonization are assumed to follow a blackbody distribution, with the thermal emission originating from the accretion disk, which is known as the eastern model. This model typically assumes a shell-like corona surrounding the NS, as most of the seed photons from the NS get scattered and Comptonized by the hot corona, while almost the entire disk can be directly observed. In Model 2, the seed photons are assumed to follow a multi-temperature disk blackbody distribution, with the thermal emission originating from the BL, which is known as the western model. The slab-like corona can be considered as representative of the western model because most of the disk photons are scattered by the corona,
while the NS radiation can be directly detected.

It is important to note that the \texttt{thcomp} model itself does not inherently allow for switching between shell-like and slab-like configurations of the Comptonization region. Instead, these geometrical interpretations are based on the assumptions made about the origin of the seed photons and the thermal component in the context of the applied models.

The best-fitting values for Model 1 are given in Table~\ref{table1}. The spectra and residuals are shown in Fig.~\ref{fig3}. Initially, we assumed a 10 kpc distance to estimate the blackbody radius ($R_{\rm{bb}}$), but it was too high (>20 km) to be consistent with a NS radius. We then adjusted to 7 kpc for more physical results. The spectral fitting results show that, as the source evolves from BOX1 to BOX3 along the Z track, the contribution from the Comptonized emission slightly increases, with the flux ratio increasing from 71\% to 84\%. In addition to the photon flux, the covering fraction of the Comptonization medium relative to the input seed photons, denoted as $f$, decreases from 0.7 to 0.39 in the HB and approaches zero in the NB. Given that $f$ approaches zero in the NB, the spectral model in the NB can be approximated as a combination of \texttt{bbodyrad} and \texttt{diskbb}. Similar findings have also been observed in the Z source GX 5--1. \citet{2024A&A...684A.137F} report that in GX 5--1, 
$f$ is approximately 1 in the HB and near zero in the NB-FB, with the PD in the HB being higher than in the NB-FB. Our analysis also reveals a distinct cooling trend in the electron temperature, decreasing from approximately 3.5 keV to 2.5 keV, along with spectral hardening as the photon index, \(\Gamma\), shifts from around 2.16 to 1.2. This behavior deviates from the typical pattern observed during state transitions in X-ray binaries. By setting \(\Gamma\) to a negative value in the \texttt{thcomp} model, we calculated the optical depth, which increased from 8.1 to 13.1 from BOX1 to BOX2. The optical depth for the NB could not be constrained by the model, possibly due to the inadequacy of the spherical geometry assumption in \texttt{thcomp} when the covering factor is extremely low. The observed increase in optical depth in the HB provides a plausible explanation for the hardening of the Comptonized component, despite the reduction in the electron temperature.

For Model 2, despite it being statistically acceptable as well, we encountered challenges in obtaining a physically meaningful solution. Specifically, the model fitting yielded \( kT_{\text{in}} \gtrsim 2 kT_{\text{bb}} \), which is physically implausible and likely due to a flat \(\chi^2\) space in the parameter region. Despite our efforts, we were unable to identify a statistically equivalent and more physically plausible solution where \( kT_{\text{bb}} > kT_{\text{in}} \). Consequently, we have decided not to present the fitting results for this model.

\section{Discussion}\label{sec4}

In this paper, we present a comprehensive X-ray spectro-polarimetric analysis of NS-LMXB XTE J1701--462, using simultaneous observations from \emph{IXPE}, \emph{Insight}-HXMT, and \emph{NuSTAR} for the 2022 outburst. During the observations, the source evolves from the HB (Epoch1) to the NB (Epoch2), tracing a Z track on the CCD. This work is focused on investigating the evolution of correlated polarimetric and spectral properties along the Z track in XTE J1701--462. In particular, the broadband energy spectra from \emph{Insight}-HXMT and \emph{NuSTAR} allow us to establish more precise constraints on the spectral parameters.

Significant polarization is detected in the HB and only a very weak polarization is detected in the NB. The PD is found to decrease from $\sim 5\%$ to $\sim 3.8\%$ along the HB in 2--8 keV, from BOX1 to BOX2, and to drop to $\leq 1\%$ in the NB (BOX3), while the corresponding PA increases along the HB and then decreases in the NB. 
The evolution of PA can be attributed to the differing PAs of the disk and Comptonized spectral components, which are neither aligned nor orthogonal. As the relative flux contributions from these components vary across different positions in the HB, the overall PA shifts accordingly. Theoretically, for axisymmetric accretion geometries, the PAs of the BL and the disk are predicted to be orthogonal to each other, allowing only 90-degree rotations in the overall PA. However, relativistic effects may cause deviations from this orthogonality, leading to changes in the overall PA as the relative flux contributions of these components evolve \citep{2022A&A...660A..25L}.

The high PD detected in the HB of XTE J1701--462 in 2--8 keV excludes the possibility of the accretion disk being the main source of the polarized radiation, since the disk emission dominates below 3 keV and the PD measured from the disk is less than 1.6\% in this source \citep{2023A&A...674L..10C}. According to the polarization measurements by \citet{2023MNRAS.525.4657J}, the increase in PD with energy further contradicts the hypothesis that the polarization primarily originates from the accretion disk.

Seed photons can become polarized after the inverse Compton scattering in the corona, which is considered one of the main sources of photon polarization \citep{1996ApJ...470..249P}. In this scenario, the PD highly depends on the geometrical configurations of accretion flow and the spectral states of the source. Specifically, for a slab-like corona, the up-scattered seed photons mainly come from the accretion disk, while for a shell-like corona, the up-scattered seed photons mainly come from the surface of the NS.

In light of these scenarios, we conducted spectral fitting analyses using data from \emph{Insight}-HXMT and \emph{NuSTAR}. The two models applied represent distinct origins of seed photons under varying coronal geometries. However, only Model 1 yielded a physically meaningful solution. The results from Model 1 demonstrate a significant redistribution between the thermal and Comptonized components along the Z track, accompanied by a notable evolution in the covering factor of the Comptonized component.

The results indicate that from BOX1 to BOX3 the flux of the Comptonized component gradually increases. If the polarization originates from inverse Compton scattering in the corona, we would generally expect an increasing PD. However, the observed PD decreases during this process. This discrepancy is likely due to the dilution of the Comptonized flux caused by a decreasing covering factor. The observed covering factor shows a decreasing trend, consistent with the PD from BOX1 to BOX3. Under the eastern model assumption, this might suggest that as the source evolves along the Z track from the HB to the NB, there is a reduction in the coverage of the BL on the NS surface.
Previous studies using Rossi-XTE observations on XTE J1701--462 have also suggested that the Comptonized emission decreases at a roughly constant accretion rate along the HB \citep{2009ApJ...696.1257L, Li:2014pza}. The synergistic evolution of PD and Comptonized radiation suggests a Comptonization origin for polarized photons. However, both \citet{2024A&A...684A..62F} and \citet{2024arXiv240916023B} have demonstrated that the PD of the BL does not exceed approximately 1.5\%, which is significantly lower than the nearly 5\% PD observed in the HB in XTE J1701--462.

The reflection component from the ionized inner accretion disk can potentially contribute up to 6\% polarization in the total spectrum \citep{1985MNRAS.217..291L}, making it a strong candidate for the polarization source. 
If polarization arises from reflection off the inner disk, the decrease in PD could be influenced by multiple factors. One of the possible reasons is a decrease in the illumination of the Compton component, which could reduce the reflection intensity. Another explanation could be a reduction in the covering factor, suggesting a reduced latitudinal extent of the BL, which limits its ability to effectively illuminate the disk. Specifically, in the NB, the gap between the inner disk and the NS could be filled by a disk coplanar accretion flow, which is geometrically thin and then poorly illuminated by the NS, as is inferred from the very low covering factor. Furthermore, as is shown in Table~\ref{table1}, the inner disk's radius significantly increases from the HB to the NB, potentially weakening reflection and causing a noticeable decrease in PD in the NB.

\section{Summary}\label{sec5}

In this study, we present a systematic analysis of the spectro-polarimetric evolution along the Z track of the NS-LMXB XTE J1701--462. Our findings reveal that the PD exhibits a monotonically decreasing trend from the HB to the NB. The broadband spectral fitting results indicate significant changes in the covering factor of the Comptonized component. These combined spectral and polarimetric measurements support a reflection origin for the observed polarization. The evolution of PD is likely driven by variations in the radiation intensity and the coverage of the BL. The recession of the accretion disk and the near-disappearance of the Comptonized emission can effectively explain the significant decrease in PD observed in the NB.

\begin{acknowledgements}

This work made use of data from the \emph{Insight}-HXMT mission, a project funded by the China National Space Administration (CNSA) and the Chinese Academy of Sciences (CAS), as well as data and/or software provided by the High Energy Astrophysics Science Archive Research Center (HEASARC), a service of the Astrophysics Science Division at NASA/GSFC.

The research was supported by the National Key Research and Development Program of China (2021YFA0718500) and the National Natural Science Foundation of China (NSFC) under grant numbers U1838202, 12273030, 11733009, 11673023, U1938102, U2038104, U2031205, 12233002, 12133007, and 12333007. Additional support was provided by the CAS Pioneer Hundred Talent Program (grant No. Y8291130K2) and the Scientific and Technological Innovation Project of IHEP (grant No. Y7515570U1). This work was also partially supported by the International Partnership Program of CAS (grant No. 113111KYSB20190020).

LD acknowledges support from the Deutsche Forschungsgemeinschaft (DFG, German Research Foundation) under Project Number 549824807.

We would also like to express our gratitude to the referee for their valuable feedback and suggestions, which have significantly enhanced the quality of this paper.

\end{acknowledgements}

%
%

\bibliographystyle{aa} 
\bibliography{aa} 
\end{document}